\documentclass{aa}  

\usepackage{graphicx}
\usepackage{txfonts}
\usepackage{xcolor}
\usepackage{amsbsy,amsmath}
\usepackage{amsfonts}
\usepackage{natbib}
\usepackage[colorlinks,allcolors=blue]{hyperref}

\begin{document}

\title{Breaking long-period resonance chains with stellar flybys}

\author{C. Charalambous \inst{1,2}
    \and
    N. Cuello \inst{3}
    \and
    C. Petrovich \inst{4}
}

\institute{Instituto de Astrofísica, Pontificia Universidad Católica de Chile, Av. Vicuña Mackenna 4860, 782-0436 Macul, Santiago, Chile \\
        \email{ccharalambous@uc.cl} 
        \and
        Millennium Institute for Astrophysics MAS, Nuncio Monse\~nor Sotero Sanz 100, Providencia, Santiago, Chile
        \and
        Université Grenoble Alpes, CNRS, IPAG, 38000 Grenoble, France 
        \and
        Department of Astronomy, Indiana University, Bloomington, IN 47405, USA
             }

\date{Received XXX; accepted YYY}

\abstract 
{Planetary migration models predict multiple planets captured into a chain of mean-motion resonances during the disk phase. Over a dozen systems have been observed in these configurations, nearly all close-in planets, with a lack of resonant chains for planets with orbital periods larger than $\sim$300 days.
}
{Dynamical studies often overlook the fact that stars do not evolve in isolation. In this work, we explore the possibility that the absence of giant planets in long-period resonant chains may be due to post-formation disruption caused by stellar flybys.}
{For planets in the 2:1-2:1 and 3:2-3:2 resonant chains, we evaluate the long-term stability after varying parameters such as the planet masses, as well as the inclination, pericentric distance, and mass of the flyby star. }
{Our integrations show that the 2:1-2:1 resonant chain is significantly more resilient to a stellar flyby than for the 3:2-3:2 configuration. The nature of the instability is different in both scenarios, the 2:1-2:1 becomes unstable quickly, soon after a penetrative close encounter. Instead, planets in the 3:2-3:2 chain become unstable in long timescales due to more distant flybys (up to $q/a_{\rm out}\sim25$ for Jupiter-mass planets) that only provide small perturbations for the system to chaotically dissolve.}
{If an encounter occurs between a star hosting planets and a passing star, Jupiter-mass systems with 3 planets in a 3:2-3:2 resonant chain or more compact initial configurations are likely to be disrupted. } 


\maketitle


\section{Introduction} \label{sec:intro}

\subsection{Resonant chains in exoplanet systems}

Planetary systems form within circumstellar disks composed mainly of gas and a minor fraction of dust. Gravitational interactions between the gaseous disk and embedded planets drive planetary migration. The direction and speed of this migration depend on the planet's mass and the local properties of the gaseous disk \citep{1979ApJ...233..857G,1986ApJ...309..846L}.

When the relative migration rate is  convergent and adiabatic (with the orbital period ratio decreasing over a timescale longer than the resonant libration period), planets in low-eccentricity orbits are expected to become trapped in mean-motion resonances (MMRs) \citep{2014prpl.conf..667B,2018haex.bookE.139N}. When more than two planets are formed in the same system, a resonant chain is expected to form, capturing all adjacent pairs in resonance.\footnote{In the 3-planet case, a mean motion resonance can be described as $k_1 n_1 + k_2 n_2 + k_3 n_3 \simeq 0$, with $k_i \in \mathbb{Z}$, and $n_i$ their mean motions. The system is actually in resonance when at least one of the associated resonant angles given by $\phi_{3{\rm pl}} = k_1 \lambda_1 - k_2 \lambda_2 + k_3\lambda_3$ librate around a fixed value. $q_{\rm 3pl} = |k_1 + k_2 + k_3|$ is called the order of the resonance \citep[see][for details on the resonance structure]{2018MNRAS.477.1414C,2021CeMDA.133...39P}. 
Resonant systems are protected from disruption and tend to be more stable than systems that are out of resonance.}

\begin{figure}
    \centering
    \includegraphics[width=\columnwidth]{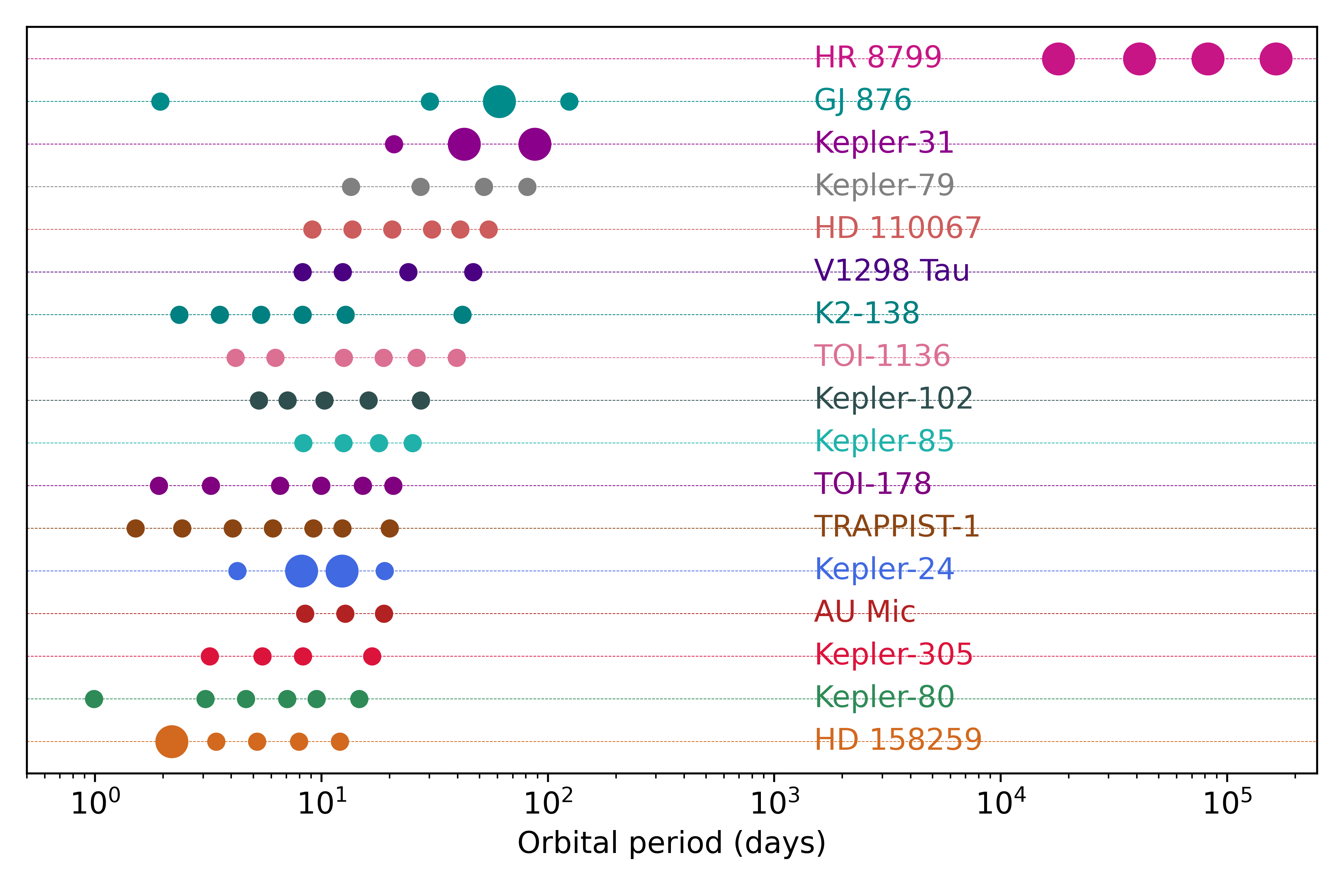}
    \caption{Systems within 5\% of resonant chains involving  the 3:2 and 2:1 period ratio commensurabilities between adjacent planet pairs. Data taken from the Catalogue of Exoplanets (\href{https://exoplanet.eu/catalog/}{https://exoplanet.eu/catalog/}).
    The size of the dots indicates planetary masses: large dots represent planets with masses of at least $1 \, m_{\rm Jup}$, while smaller dots correspond to less massive planets. }
    \label{fig:observed_chains}
\end{figure}
However, out of $\sim$5500 known planetary systems, around 1000 host multiple planets, yet fewer than 30 exhibit resonant chains. Figure \ref{fig:observed_chains} illustrates systems near first-order resonances (only those within 5\% of period ratios implying 2:1 and 3:2 MMRs), based on data retrieved from the Exoplanet Catalogue (\href{https://exoplanet.eu/catalog/}{https://exoplanet.eu/} as of November 2024). 

The data reveals that, with the exception of HR~8799, all systems exhibiting resonant chains consist of super-Earths with orbital periods typically within 200 days of their host star.
In contrast, cold Jupiter systems—a population of giant planets with semi-major axes beyond 1 au— are frequently observed in isolation \citep[or at least, lacking close companions, e.g, with period ratios below $\sim5$,][]{zhu2022,rosenthal2024}. Their significant eccentricities (median of $\sim 0.2$, e.g., \citealt{kipping2013,2020AJ....159...63B,2021ApJS..255....8R}) largely suggests that planetary ejections from planet-planet scattering has shaped their orbits and decreased their multiplicities \citep[see][and references therein]{1996Natur.384..619W,2008ApJ...686..580C,JT2008,2014prpl.conf..787D}.

Yet, a small fraction ($\sim 15\%$) of these cold Jupiters are observed alongside with other cold Jupiter companions, frequently observed near the 2:1 MMR and to a lesser extent near the 3:2 \citep{2022MNRAS.514.3844C, 2022MNRAS.513..541C}. This leads to the following central questions: what mechanisms can disrupt long-period resonant chains of systems with giant planets? Could isolated cold Jupiters result from the breakup of such chains?

For close-in super-Earth planets, several processes have been identified that prevent the formation of resonant pairs, such as overstable librations \citep{2014AJ....147...32G}, disk turbulence \citep{2017AJ....153..120B}, planetesimal-driven migration \citep{2008ApJ...686..580C}, and tidally-driven resonance divergence induced by the disk \citep{2020MNRAS.495.4192C} or tides acting on the planets \citep{2012ApJ...756L..11L,2013AJ....145....1B,2023A&A...677A.160C}. Recently, \citet{2024arXiv240606885D} showed that resonant configurations are practically universal among young planetary systems. Over time, however, these resonances tend to dissolve. This supports the idea that planets are initially assembled in resonant configurations, which are gradually disrupted by various processes, including dynamical instabilities such as those suggested by the `breaking-the-chains' model \citep{2017MNRAS.470.1750I,2022AJ....163..201G}. 

For cold Jupiters, the mechanisms responsible for breaking MMRs remain more elusive. Tides are inefficient \citep{2012ApJ...756L..11L,2013AJ....145....1B}, and planetesimal-driven migration and turbulence are less effective due to the higher planetary masses \citep{2008ApJ...686..580C}.
One possibility is that planets never got captured in the first place. However, the recent numerical experiments by \citet{nagpal2024} show that the conditions to disrupt the resonances for giant planets are somewhat stringent, demanding fast migration rates and inefficient eccentricity damping rates\footnote{Values of $K \equiv\tau_a /\tau_e\sim 1-10$, where $\tau_a$ and $\tau_e$ are the radial and eccentricity damping timescales, respectively, for the migrating planets.}--conditions not naturally accommodated within the type-II migration paradigm. 
Furthermore, in systems that still retain a disk, such as PDS 70 and HD 163296, observational evidence suggests that long-period resonant chains can form \citep{2019ApJ...884L..41B,2023ApJ...945L..37G}, while young systems like HR~8799 still preserve such resonance structure \citep{2020ApJ...902L..40G,2022A&A...666A.133Z}. As such, resonance chains for cold Jupiters may indeed be broken after the disk dispersal phase by a process that has yet to be determined.

In this work, we propose a mechanism analogous to the breaking-the-chains model for gas giant planets, where resonant chains are formed through disk-driven migration. Our key contribution is to consider the system as non-isolated, evaluating the role of stellar flybys in breaking these resonances.

\subsection{Planetary systems perturbed by flybys}

Most models assume that stars are isolated and that the process of planet formation and evolution occurs almost unperturbed. However, it is well known that following the collapse of molecular clouds, stars often form multiple stellar systems \citep{Pringle1989, Reipurth+2014, Offner+2023}, which evolve within stellar clusters and associations \citep{LadaLada2003}. In these regions, stellar encounters are not only possible but rather frequent. The likelihood of these events strongly depends on the adopted flyby definition and it increases with higher stellar density \citep{2013A&A...549A..82P, 2018MNRAS.478.2700W}, as expected. Here, we follow the typical flyby definition formulated in \cite{Cuello+2023}: a stellar encounter occurs whenever two stars --- with eccentricity equal to or greater than 1 --- get closer than 1000~au. For instance, for a 3~Myr old Solar-like star in a leaky cluster or in an OB association, the probability of experiencing a flyby is around 20\% according to \cite{2013A&A...549A..82P}. By that time, the process of planet formation should already have taken place \citep{Manara+2023}, meaning that stellar flybys can impact fully-formed planetary systems, which may already be in resonance.

Scenarios where a mature planetary system is perturbed by an intruding star have been studied over the years. \citet{1998ApJ...508L.171L} showed that stellar encounters in dense open clusters can cause significant orbital disruption, which could explain the presence of Jovian planets in eccentric and relatively close orbits. In a similar vein, \citet{2011MNRAS.411..859M} explored through numerical simulations how stellar flybys can trigger scattering events among planets, disrupting stable configurations. \citet{2015MNRAS.448..344L} compare the effects of single and binary stars, finding that binaries generally have a greater impact on planetary systems, particularly those with widely spaced planets. \citet{2019MNRAS.488.1366L, 2020MNRAS.496.1149L} further explore scenarios in which planetary systems lose planets due to catastrophic flybys, where the passing star, that has its own planetary system, is allowed to cross between the planets of the affected system, significantly disrupting their orbits.

Our approach is novel since it focuses on mature systems with resonant chains in wide orbits, while previous studies examined compact systems in their birth clusters. Here, we explore systems in the field with giant planets perturbed by a single passing star, which, despite fast relative velocities, has a non-negligible impact \citep{2012MNRAS.425..680V}. Additionally, we analyze the effects of the passing star at realistic distances in different resonant chains and assess their long-term fates. The remainder of this paper is organized as follows. Section \ref{sec:simus} outlines the typical flyby parameters, and details the setup for the simulations. In Section \ref{sect:results}, we discuss the results of our simulations after varying several parameters, and in Section \ref{sect:conclusions} we summarize our findings.

\section{Flyby parameters and setup of the simulations}
\label{sec:simus} 

We consider systems of three planets on coplanar orbits around a central star with mass $M_\star = 1 \, M_\odot$. We denote the inner, middle, and outer planetary masses as $m_1$, $m_2$, and $m_3$, respectively, with $M_\star \gg m_1, m_2, m_3$. The three planets in the system are interconnected through resonant chains: planets 1 and 2 are in a mean-motion resonance, as are planets 2 and 3. Consequently, the entire system exhibits a three-planet resonance, a multi-resonant state. Our focus is primarily on zero-order 3-planet resonances ($q_{3 \rm{pl}}=0$, also known as Laplace resonance), as these are the only ones observed among exoplanets at present\footnote{In the case of the system K2-138, there is a claim that the outermost planet may not be decoupled from the inner system as previously thought. Instead, it might follow a first-order three-planet commensurability \citep[see][]{2023ApJ...954...57C}, although this has yet to be confirmed observationally.}.

Since we are interested on the effects of the flyby on the system's dynamics, we force the planets into a resonant chain, mimicking in situ formation and only experiencing short-term migration. We adiabatically push the outer planet inwards, allowing it to capture the inner planets in resonance, leading to the libration of the two- and three-planet resonant angles. Orbital and eccentricity damping timescales of $\tau_a=10^6$ yrs and $\tau_a/\tau_e=100$ are applied to account for planet-disk interactions, with the planets initially placed 10\% away from resonance. The capture takes place at the beginning of the simulation, after which we let the system evolve in isolation for $5\times 10^5$ yrs, influenced only by their gravitational interactions to ensure stability. After this period, a passing star with a mass of $M_{\rm fb} = 1\, M_\odot$ approaches the system. It is important to mention that in the absence of a stellar flyby, both the 2:1-2:1 and 3:2-3:2 configurations remain stable and locked in resonance for at least $10^7$ orbits of the innermost planet, $\sim 5\times 10^8$ years, with typical equilibrium eccentricities lower than $5\%$, and the libration amplitude of the resonant angles around $20^\circ$.

We consider that a flyby occurs whenever an unbound star from the field gets closer than 1000 au to the planet-hosting star \citep{Cuello+2023}. In order to avoid spurious gravitational effects on the planetary system under study and following \cite{Bhandare+2016}, we initialize the perturber star at 20 times the distance of minimum approach (also referred to as orbital pericenter). \cite{Breslau+2017} showed that considering different azimuthal positions for a given planet when the perturber goes through pericenter leads to different dynamical outcomes (perturbation, ejection or capture). To include this effect, for a given numerical setup, we perform eight 45-degree rotations of the initial true anomaly (position along the orbit). This approach allows us to obtain meaningful results since, for a given resonant layout, we take into account how different initial conditions affect the dynamical outcome.

To sample a limited but representative region of the parameter space, we consider individual stellar perturbers with masses and pericenter distances ranging from 0.5 to 2 $M_\odot$ and from 75 to 500 au, respectively. These values are consistent with recent calculations of stellar cluster dynamics \citep{Schoettler+2024}. For instance, \cite{Schoettler+2020} find that more than half of the encounters occur at distances closer than 300 au in the Orion Nebula Cluster. In more sparse environments, stellar encounters still occur at a moderate rate with low velocities and at pericenter distances of several hundred au \citep{2013A&A...549A..82P, 2018MNRAS.478.2700W, 2024A&A...691A..43W}. It is worth noting that the likelihood of an encounter decreases with the cluster age. However, as will be shown in this work, under certain conditions, a distant flyby is enough to disrupt a resonant chain of planets. If the system would contain gas, the perturbative effects would be damped \citep[e.g.][]{Marzari+2013, Nealon+2020}. This is why we decide to focus on gas-less planetary systems. Therefore, the effects reported here should be considered as upper limits for planetary systems embedded in gaseous disks. In addition, as done in \citet{Cuello+2019}, we assume that there is no preferential direction in which the flyby occurs, and we therefore consider 5 possible orbital inclinations with respect to the initial planetary system orbital mid plane: i) coplanar prograde, ii) inclined prograde, iii) polar or orthogonal, iv) inclined retrograde, and v) coplanar retrograde.

In this scenario, the velocity at which the perturber crosses the sphere of 1000 au centered around the planet-hosting star determines the type of encounter. For very high speeds, one can use the impulse approximation assuming that the planets do not progress in their orbits around the central star during the encounter. Typically, these flybys are penetrating or catastrophic. On the contrary, if the perturber has a large pericenter distance and arrives at a lower speed, then we can describe the perturbation in the secular regime as the planets complete various orbits during the interaction. In such case, the flyby is expected to lead to variations in the orbital eccentricity while the planetary semi-major axis should remain almost unchanged (unless the perturber crosses the planetary orbits on a parabolic orbit).

Taking into account all the points mentioned above, our fiducial simulation involves a system consisting in three Jovian planets with $(0.75, 1, 1.25) \, m_{\rm Jup}$ in a resonant chain, and a 1 Solar mass star approaching the planetary system with different inclinations $ \beta \in [0^\circ, 45^\circ, 90^\circ, 135^\circ, 180^\circ]$. For the minimum distance between the stars, we consider six cases: $q \in [75, 100, 125, 150, 175, 200]$ au. We analyze two specific first-order resonance chains, namely the 3:2-3:2, and 2:1-2:1. In each setup, the outer planet is positioned at $\sim 40$ au, ensuring that the geometric cross section of both layouts remains the same. With these parameters, the encounter can be regarded as a secular perturbation \citep[see][]{1996MNRAS.282.1064H,2009ApJ...697..458S}. For each combination, we perform eight rotations per system modifying the angular position of the planets along their respective orbits, resulting in a total of 480 simulations. 
We perform our simulations using the N-body code \textsc{REBOUND} \citep{2012A&A...537A.128R} employing the IAS15 integrator \citep{2015MNRAS.446.1424R}, incorporating a stellar flyby on a parabolic orbit. Once the flyby is over (i.e. when the perturber is at more than $1000$ au from the primary star), we remove the perturber from the simulation. Finally, we evolve the planetary system for another $10^5$ orbits of the innermost planet, using the hybrid integrator TRACE \citep{2024MNRAS.533.3708L}. TRACE uses the symplectic WHFast integrator \citep{2015MNRAS.452..376R} when particles are well separated, and automatically switches over to a Bulirsch-Stoer/IAS15 integrator during close encounters.

\section{Results} \label{sect:results}

\begin{figure*}
    \centering
    \includegraphics[width=\textwidth]{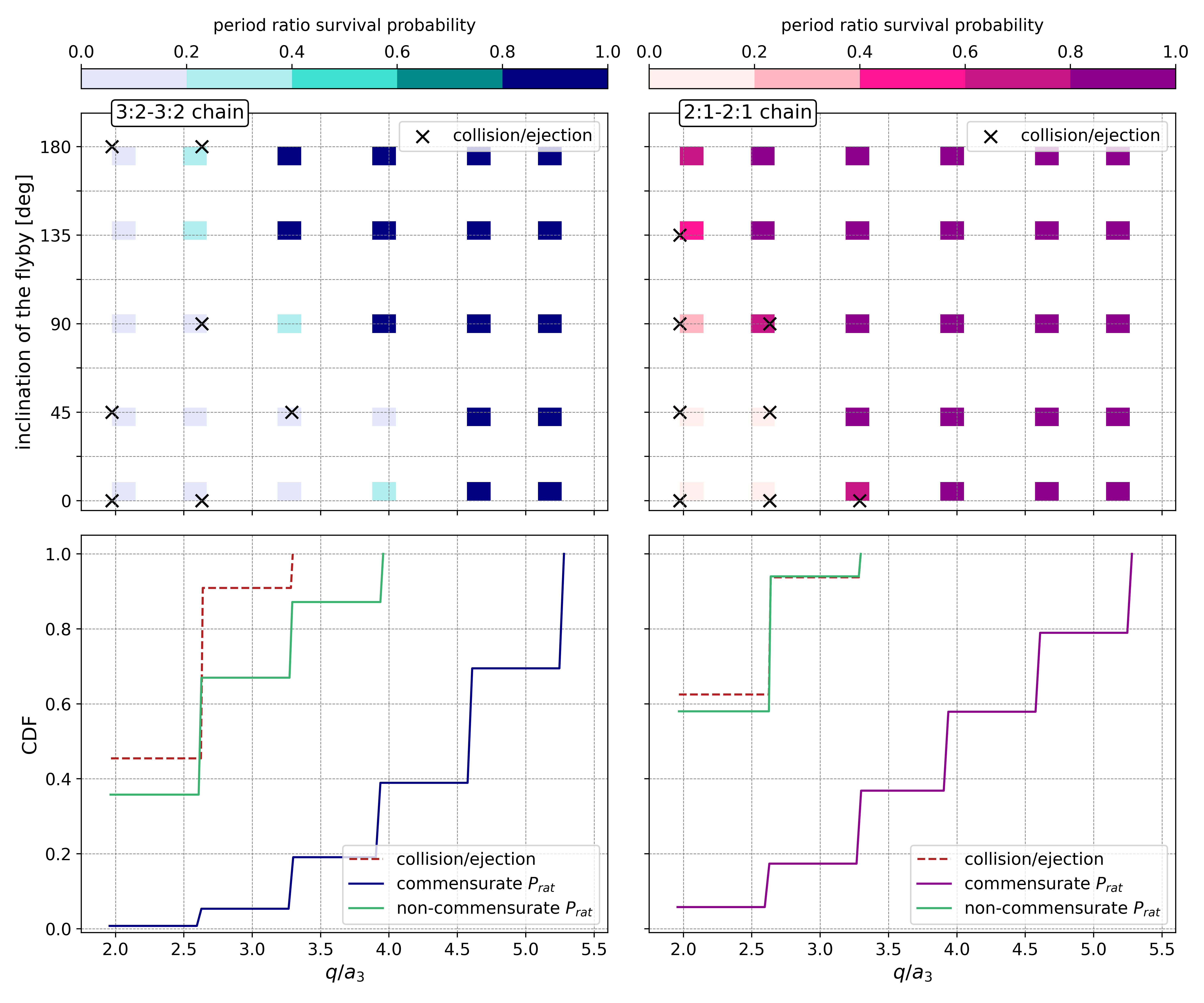}
    \caption{Top: Survival probability of period ratios (shading from lighter to darker) immediately after a star with a mass of $M_{\rm fb} = 1 \, M_\odot$ flies by a three-planet system with $(0.75, 1, 1.25)$ Jupiter masses in 3:2-3:2 (left) and 2:1-2:1 (right) resonant chains and leaves the surroundings of the system to a distance $>1000$ au, considering different inclinations and distances relative to the orbital plane of the planets. 
    Bottom: Cumulative Distribution Function (CDF) for the possible outcomes of the simulations: collisions with the star and ejections (red dashed line), survival of the period ratios (their resonant angles not necessarily librating, in solid purple), systems where period ratios do not remain commensurate (green), plotted against the pericentric distance of the flyby star relative to the outermost planet. }
    \label{fig:q_i_chains}
\end{figure*}

Given that giant planets on wide orbits around solar type stars may not be uncommon \citep{fernandes2019,fulton2021}, our working hypothesis is that these planets initially formed in resonant chains on wide orbits. In this section, we examine the possibility that the system maintains its resonant configuration after an interaction with a nearby star. 
Section \ref{sect:different_resonances} focuses on the evolution within the 2:1-2:1 and 3:2-3:2 resonant chains, while Sections \ref{sect:planet_mass} and \ref{sect:star_mass} discuss the impact of planetary and stellar masses on these interactions.

\subsection{Effects of a stellar fly-by on wide orbit resonant chains}\label{sect:different_resonances}
\subsubsection{The immediate response}
\begin{figure*}
    \centering
    \includegraphics[width=\textwidth]{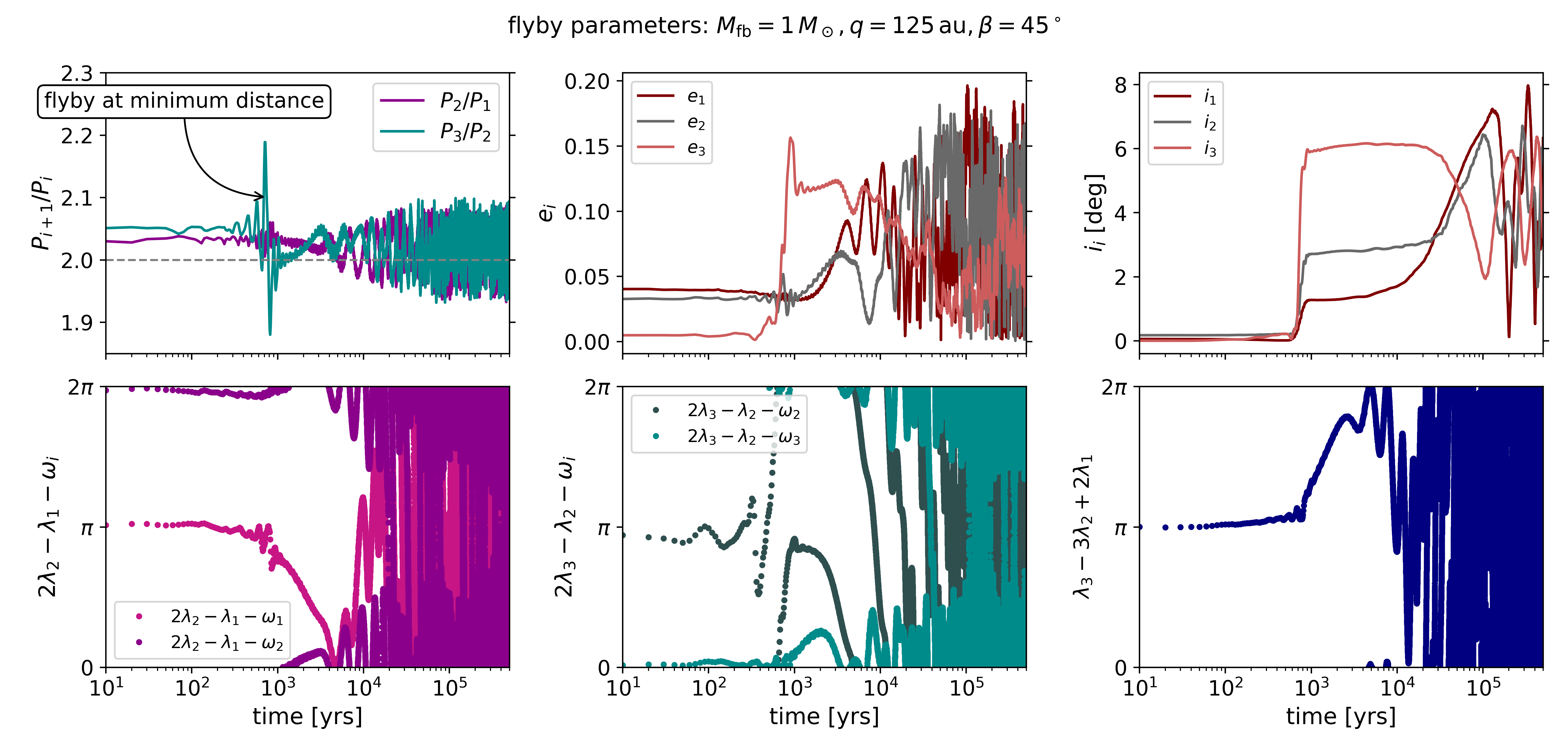}
    \caption{Typical time evolution for the three planets in the 2:1-2:1 resonant chain after a close encounter with a flying-by star with mass $1\, M_\odot$, and inclination $\beta=45^\circ$. The distance of minimum approach between the two stars is $q=125$ au. In the top panels from left to right, we show the period ratio between the adjacent pairs, the eccentricities and inclinations, while in the bottom panels, the two-body resonant angles for each pair (left and middle plots) and the three-body resonant angle (in the bottom right panel). }
    \label{fig:evol}
\end{figure*}
\begin{figure*}
    \centering
    \includegraphics[width=\textwidth]{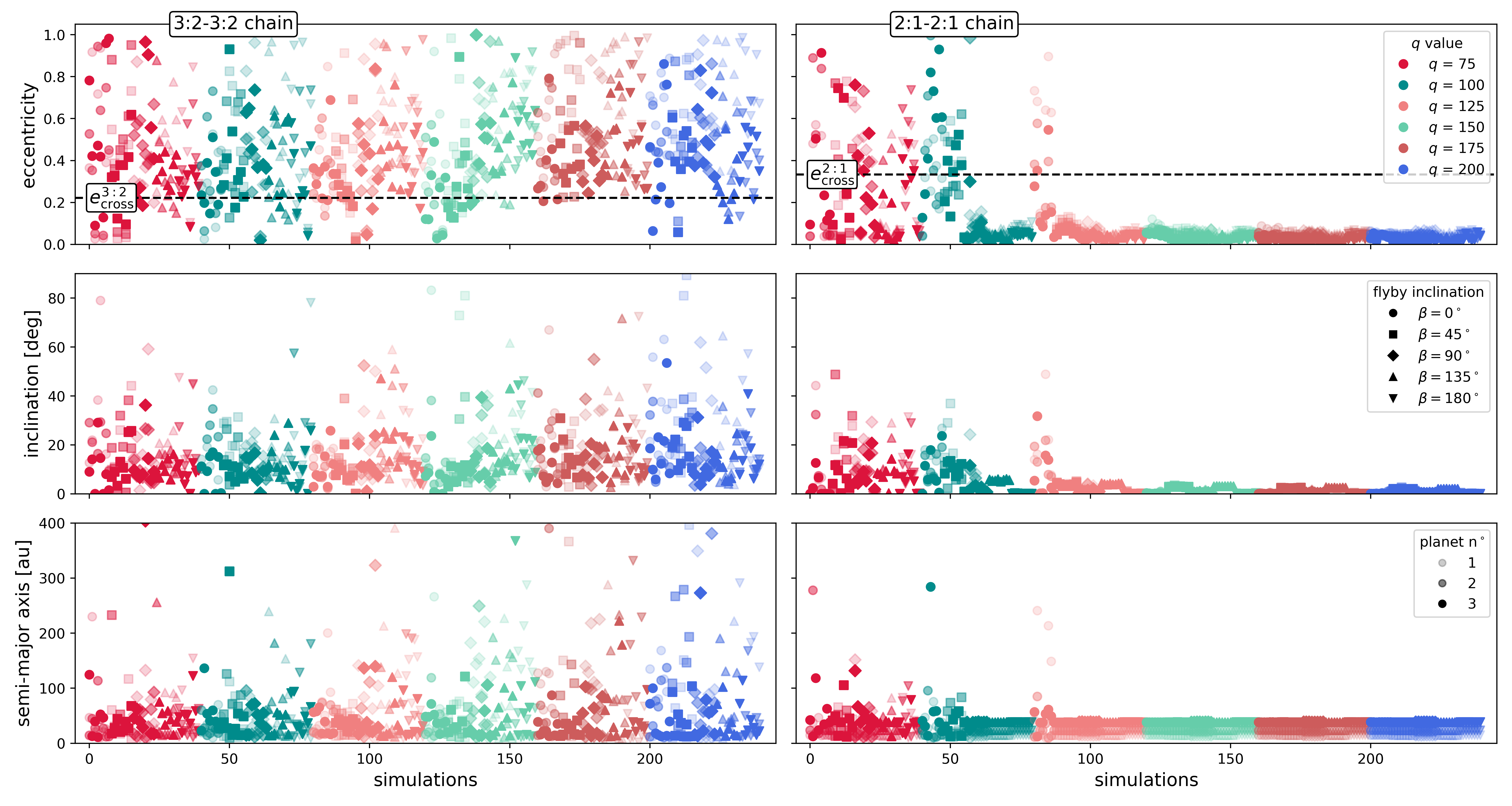}
    \caption{Delayed eccentricities (top row), inclinations (middle) and semi-major axis (bottom row) 
    for all simulations in the 3:2-3:2 (left column) and 2:1-2:1 (right column) resonant chains. Colors indicate the pericentric distance $q$ between the stars during the flyby event, with each color corresponding to a specific $q$, while the symbols represent the inclination $\beta$ of the flyby star. The black dashed line indicates the stability limit given by the orbit-crossing value $e_{\rm cross}$. }
    \label{fig:e_i_sims}
\end{figure*}
\begin{figure}
    \centering
    \includegraphics[width=\columnwidth]{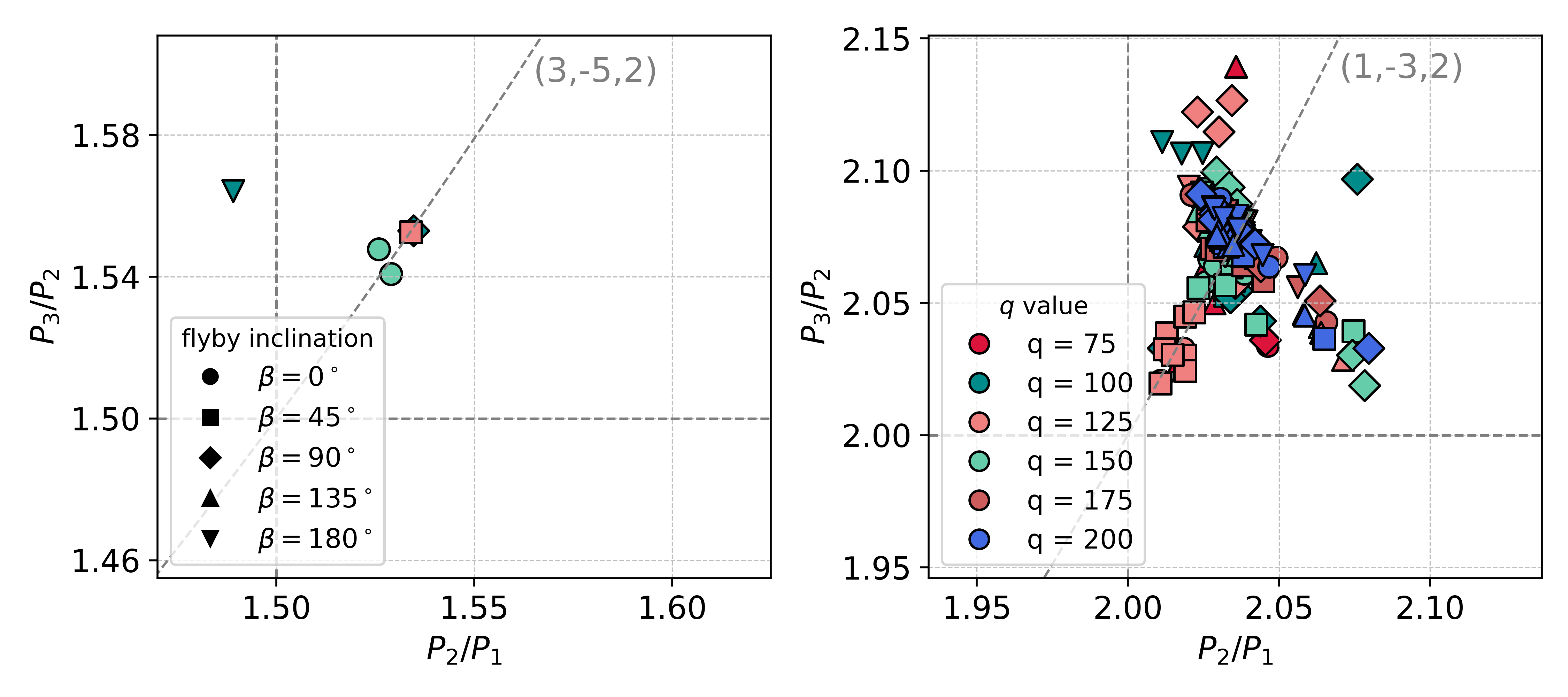}
    \caption{Conditions leading to configurations within 5\% of the 3:2-3:2 (left) and 2:1-2:1 (right) resonant chains in the period-ratio plane $(P_2/P_1, P_3/P_2)$. Relevant two- and three-planet MMRs are indicated by dashed gray curves. Same colors and markers as in Fig.~\ref{fig:e_i_sims}. }
    \label{fig:Prat}
\end{figure}

In the top panels of Figure \ref{fig:q_i_chains}, we present the outcomes of our simulations for systems initially trapped in a 3:2-3:2 (left) and a 2:1-2:1 resonant chain (right). The color coding represents the survival probability of period ratios, with darker colors indicating higher probabilities and lighter colors indicating lower probabilities. Crosses denote instances of ejections or collisions with the star, resulting from destabilization due to the flyby. The $y$-axis in both panels shows the inclination of the passing star, while the $x$-axis shows the distance of minimum approach normalized by the position of the last planet in the system ($40$ au in all cases). 

We assess the systems' stability by considering whether they maintain their orbital periods within 5\% of the fixed point after the passing star moves beyond a distance of 1000 au from the central star, a criterion we refer to as the immediate response. Although the outer planet in both scenarios is placed at the same distance and, therefore, similarly perturbed by the passing star, Figure \ref{fig:q_i_chains} shows that the 3:2-3:2 layout is more susceptible to disruption than the 2:1-2:1. In the 3:2-3:2 setup, the planets are closer to each other when the flyby occurs. This closer proximity, combined with the compact nature of these configuration, makes the system more vulnerable to destabilization by stochastic perturbations.

When the passing star approaches the system in the same direction as the planets' orbital motion ($\beta = 0^\circ$), it disrupts the resonant chain more effectively than during a retrograde approach, even at greater encounter distances. \citet{2019MNRAS.488.1366L} observed a similar effect, where prograde encounters were more likely to destabilize the system and allow the intruding star to capture planets.
By integrating across all inclination values, we obtain the cumulative distribution distribution (CDF) of $q/a_3$ for all possible outcomes: collision/ejection, preservation of orbital periods, and overall stability, as shown in the bottom panels. The immediate response of the system to the effects of the flyby in the 2:1-2:1 resonant chain shows that whenever $q/a_3$ exceeds 3.5, the systems maintain their constant period ratios, while in the 3:2-3:2 configuration, the systems retain the resonant relation for $q/a_3>4.5$.

As mentioned in Section \ref{sec:intro}, the fact that the period ratios of consecutive pairs lay in (or close to) an integer ratio does not necessarily mean the system is resonant. For that purpose, the resonant angles must librate around a fixed value. In Figure \ref{fig:evol} we show a typical temporal evolution of the planets immediately after the close encounter with the passing star. In this example, the pericentric distance of the approaching star is $q=125$ au ($q/a_3 \sim 3.5$), with an inclination of $\beta = 45^\circ$. As expected, the outermost planet is most affected by the interaction with the stellar flyby, causing a bigger instantaneous change in $P_3/P_2$ than in $P_2/P_1$. However, while the period ratios remain locked in a 2:1-2:1 commensurability after the passage of the star, the resonant angles begin to circulate, meaning that the system has, in fact, left the resonance. 

In such cases, the systems are left with commensurate period ratios but lack the protection provided by the resonance \citep[see ][]{2008ApJ...686..580C}. The final orbital elements after the flyby differ significantly from those just after the resonant capture, which jeopardizes the system's dynamical stability. While secular timescales are orders of magnitude longer than orbital timescales, in this first set of simulations, we do not integrate over such extended periods. Instead, we keep track of the resonant angles for several libration cycles following the encounter. However, to assess the system's ability to maintain constant period ratios and evaluate their long-term outcomes, it is essential to consider the secular evolution. Therefore, we conduct additional simulations, extending the integration by $10^5$  orbits of the innermost planet.

\subsubsection{The delayed response}
Figures \ref{fig:e_i_sims} and \ref{fig:Prat} present the results of the delayed interactions between the passing star and the planets, with pericentric distance of the flyby represented in different colors. The symbols show the flyby's inclination, and the transparency of the symbols depicts the planets' positions in the system prior to the encounter. From top to bottom of Fig.~\ref{fig:e_i_sims}, we show the eccentricities, inclinations and semi-major axis, respectively. For clarity, we exclude planets that end the simulations beyond 400 au, which is 10 times the initial distance of the outermost planet. In Fig.~\ref{fig:Prat} we show the systems that survive the flyby in the period ratio plane.

For the chosen parameters, the encounter between the passing star and the planetary system is slow and tidal ($q \gg a_3$), occurring within the secular regime. For a single planet, the change in the eccentricity can then be expressed using the formulation by \citet{1996MNRAS.282.1064H} as
\begin{equation}
    \delta e = -\frac{15 \pi}{16} \left( \frac{2 M_{\rm fb}^2 a^3}{M_{123}M_{12}q^3} \right)^{1/2}
    e \sqrt{1-e^2} \sin(2\Omega) \sin^2 \beta,
\end{equation}
where $M_{12} = M_\star + m_{\rm pl}$ and $M_{123} = M_{\rm fb} + M_{12}$. Encounters at larger relative distances cause smaller changes in eccentricity, while closer encounters result in more significant perturbations. This power-law dependence on the ratio $(a/q)^{3/2}$, which we have verified for a single planet through simulations, shows that even distant stellar flybys, despite being slow, can induce long-term changes in the orbital parameters of planetary systems. The strength of the perturbation decreases with increasing distance from the passing star, making closer encounters far more disruptive to the planet's orbit. 

In the case of systems with three planets, these assumptions still hold: the encounter is slow and tidal relative to the outermost planet, and the perturbations happens in the secular limit. As shown in Figure \ref{fig:e_i_sims}, the changes in semi-major axis are less significant than those in eccentricities, which is the most perturbed orbital element. The important thing here is the amount of energy exchanged during the encounter, even very small changes in eccentricity can have profound effects, potentially leading to close encounters and, eventually, the disruption of the system. The closer the encounter, the more significant the disruptions to the orbits of any planets in the system. Such events may have played a major role in shaping the currently observed exoplanet population, particularly by contributing to the scarcity of wide-orbit systems in 3:2-3:2 resonant chains. 

In all simulations where eccentricities exceed the crossing-orbit value, $e_i > e_{\rm cross} \simeq 2k/3j$ for a $j:j-k$ MMR \citep[e.g.,][]{2021AJ....162..220T}, planets are eventually ejected, ultimately breaking the chains or leaving one planet at a distance dynamically decoupled from the inner pair. In the 3:2-3:2 chain, these changes are particularly pronounced, even during distant flybys, making it more difficult to find exoplanetary systems in this configuration on wide orbits, even if the event is not catastrophic.

\begin{figure*}
    \centering
    \includegraphics[width=\textwidth]{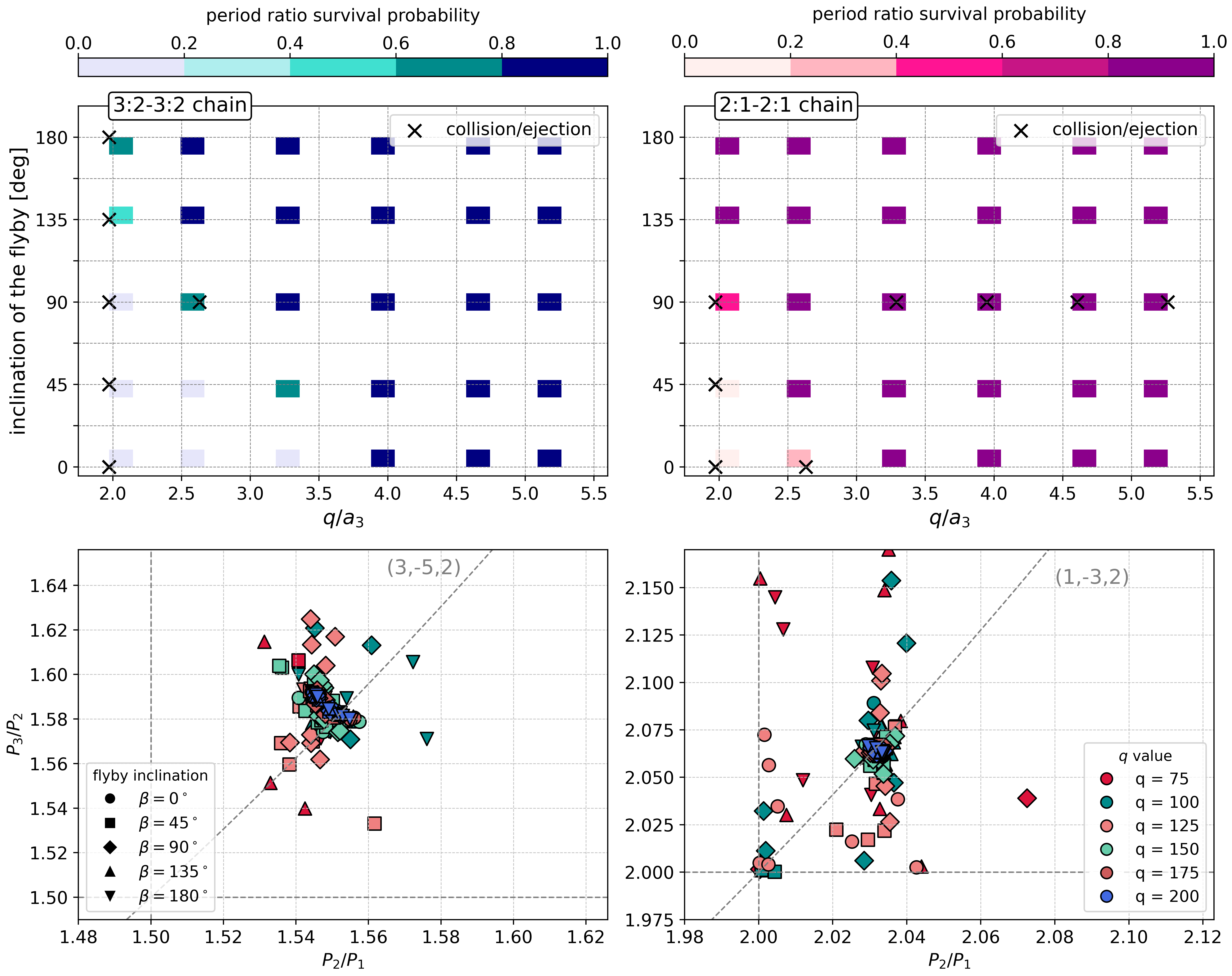}
    \caption{Top panels: Period ratio survival for sub-Saturn mass planets, in 3:2-3:2 (left) and 2:1-2:1 (right) resonant chains. Bottom panels: delayed survival (after $10^5$ orbits of the innermost planet) in the period-ratio plane. Colors and markers follow the scheme used in Figure \ref{fig:e_i_sims}. }
    \label{fig:subSat}
\end{figure*}

Scattering within the systems triggers subsequent orbital instabilities, where planets gain eccentricity, leading to crossing orbits (simulations that end above the dashed lines in the top panels). This process often results in planets either being ejected or colliding with the host star, even long after the stellar perturber has left the planetary system surroundings. Closer encounters, as well as more distant encounters in prograde orbits, induce more significant changes in the orbital elements.

It is important to note that while Figure \ref{fig:e_i_sims} does not show simulations with flyby pericentric distances beyond $q>200$ au, we conducted simulations with flybys up to $q=1000$ au.
Flybys at such large distances should exert negligible gravitational effects on planetary systems, with typical sizes between 10 and 100 au. However, in the 3:2-3:2 resonance setup, we find that although systems with $q/a_3>4.5$ ($q=175$ au) initially survive the encounter, only some remain stable over secular timescales. This results from the marginal stability of such massive planets in compact configurations, where even minor perturbations can break the system. We also observe that, although not shown here, reducing the planetary masses improves the stability of this configuration. A Jupiter-mass system in the 3:2-3:2 resonant chain is at the brink of instability, and the flyby provides just enough perturbation in the eccentricities (less than 1\%) to disrupt the stability provided by the resonance.

In contrast, nearly all period ratios in the 2:1-2:1 chain remain within 5\% of the exact integer value once the pericentric distance of the passing star exceeds $q > 125$ au (equivalently $q/a_3 \gtrsim 3.5$). However, as with the 3:2-3:2 chain, the resonant angles leave the commensurability. Although the value we obtain for $q/a_3$ is slightly smaller than that reported in \citet{2015MNRAS.448..344L}, who examined four-planet systems perturbed by a passing binary and a direct comparison is not possible, their observation that the cross-sections for planetary systems interacting with single stars are smaller than those for binary encounters suggests that our results are consistent with theirs. Besides, note that for $q=125$ au, when the flyby is coplanar (represented by circles), the planets are left on inclined orbits. This contrasts with the case of a star passing near a protoplanetary disk, where, after relaxation, the particles return to the plane of the disk \citep{Marzari+2013, Picogna+2014, Nealon+2020}.

In the cases in which the flyby disrupts the chain and the systems go unstable, the effect for the two  different configurations generates different eccentricities for the remaining planets. The eccentricities acquired by planets that survive the extended simulations are higher if the system formed near the 3:2-3:2 resonant chain compared to those originally trapped in the 2:1-2:1. 

When comparing the immediate versus delayed effects of the flyby (Figures \ref{fig:q_i_chains} and \ref{fig:e_i_sims}, respectively), we find little difference in the 2:1-2:1 resonant chain, but observe a significant difference in the 3:2-3:2 setup. Analyzing systems within the 3:2-3:2 resonance immediately after the star departs might suggest that the planets retain their period ratios. However, over time, planet-planet interactions eventually destabilize the system. As previously mentioned, although the encounter between the passing star and the planetary system is secular, the immediate effect of the flyby is most noticeable in eccentricities, which induces long-term variations in the semi-major axes. In contrast, the long-term effects on the 2:1-2:1 resonance seem to have minimal impact on the overall configuration, as shown in Figure \ref{fig:Prat}. However, while the period ratio and orbital elements remain largely unaffected, the resonant angles circulate. 

In the work of \citet{2022MNRAS.513..541C}, it is noted that large-mass systems show a strong preference for orbital period ratios close to the 2:1 commensurabilities, rather than for the 3:2 and more compact configurations. Since the long-term effects of the flyby are considerably stronger in the 3:2-3:2 resonant chain compared to the 2:1-2:1, particularly in producing substantial changes in eccentricities, it suggests that resonant chains in the 3:2-3:2 layout are unlikely to survive a flyby. Therefore, systems in this configuration are less likely to be observed, especially considering the high probability of a past close encounter between stars \citep{2013A&A...549A..82P}.

To describe the dynamics of a single zeroth-order 3-planet resonance, it is possible to write an integrable one-degree-of-freedom Hamiltonian to second order in eccentricities \citep{2011MNRAS.418.1043Q,2018MNRAS.477.1414C,2020A&A...641A.176P}. This formulation uses the 3-planet angle $\phi_{\rm 3pl}$ (see Section \ref{sec:intro}), which is independent of $\varpi_i$. As a result, the system preserves its angular momentum deficit \citep{1997A&A...317L..75L}. Consequently, if the system is only influenced by these resonances, initially circular orbits will remain circular. 

When the resonances are well isolated, there is no possibility for large-scale chaos. In such cases, the system behaves as nearly secular and can be considered as long-term stable. However, due to the perturbation introduced by the passing star, the eccentricities get excited, and the systems can no longer be considered stable (see Figure \ref{fig:e_i_sims}). In the absence of a stellar flyby, the systems in our simulations remain stable for at least $10^8$ years, with all angles librating.  However, the flyby causes the three-planet resonant angle to circulate, triggering changes in eccentricities and leading to the observed instabilities in the 3:2-3:2 resonant chain. Opposite to this, in the 2:1-2:1 configuration, even with the Laplace angle circulating, the systems are able to maintain the eccentricities low enough that instabilities do not arise.

In summary, while the immediate response of planetary systems to a nearby stellar flyby may appear similar across the configurations analyzed, the long-term outcomes differ significantly. Systems initially near a 2:1-2:1 resonant chain are more likely to retain their period ratios, implying a resonant configuration after long-term interactions, albeit often with circulating resonant angles. 
In contrast, systems starting near a 3:2-3:2 configuration experience planet-planet scattering triggered by the flyby, producing a population of eccentric cold Jupiters from orbits that would otherwise remain stable. Lastly, recent experiments by Maas et al., (\emph{in prep.}) including a sequence of flybys from cluster simulations applied to HR~8799 showed that in the absence of resonant chains (i.e., with non-librating resonant angles), the survival probability of HR~8799-like systems is nearly zero, a stark contrast to the higher stability observed in resonant systems.

\subsection{The effect of the mass of the planets} \label{sect:planet_mass}

\subsubsection{Sub-Saturn mass planets}
    
In Section \ref{sect:different_resonances}, we examined the dynamical evolution of the system post stellar flyby, focusing on scenarios with Jupiter-mass planets in different resonant configurations. In this section, we extend our analysis by investigating how varying the individual planet masses impacts the stability of the system. To achieve this, we replicate our fiducial set of simulations, but instead of Jupiter-mass planets, we consider sub-Saturn mass planets of $(40, 53.33, 71.11) \, m_\oplus$. While the mass ratios between the planets are kept constant, the ratios between the planets and the central star have changed.

In the top panels of Figure \ref{fig:subSat}, we present the immediate response of the planetary orbits to the stellar flyby. On the left, we show the case of a 3:2-3:2 resonant chain, while the right displays the 2:1-2:1 resonance. The bottom panels depict the delayed response of the system in the period ratio plane. Consistent with our fiducial Jupiter-mass simulations, a retrograde stellar flyby continues to enhance the likelihood of the planetary system's to maintain constant period ratios. Opposite to what we observe for giant planets, for systems of lower-mass planets, flybys beyond $q = 125$ au still permit the system to remain intact with period ratios within 5\% of the 3:2-3:2 configuration, at least in the considered timescales, although the resonant angles are not librating. This is in contrast to the Jupiter-mass case, where more distant flybys destabilize the system. Collisions the star and ejections happen when the star passes closer to the system compared to the case of the giants.

Additionally, for closer encounters, the final configuration of the planets shows a broader dispersion in the period ratios $P_2/P_1$ and in $P_3/P_2$, centered around the location of the three-planet MMR, particularly in the 2:1-2:1 case. This spread occurs independently of the inclination of the flyby and highlights the greater robustness of systems with smaller planets to close stellar encounters.

In short, lower-mass planets exhibit a higher degree of resilience to perturbations induced by stellar flybys compared to their giant counterparts. This suggests that a multi-resonant state involving low-mass planets can survive in systems where stellar encounters occur at relatively large distances, opening the possibility of finding such systems in wide orbits.

\subsubsection{Super-Jupiter mass planets}
\begin{figure}
    \centering
    \includegraphics[width=\columnwidth]{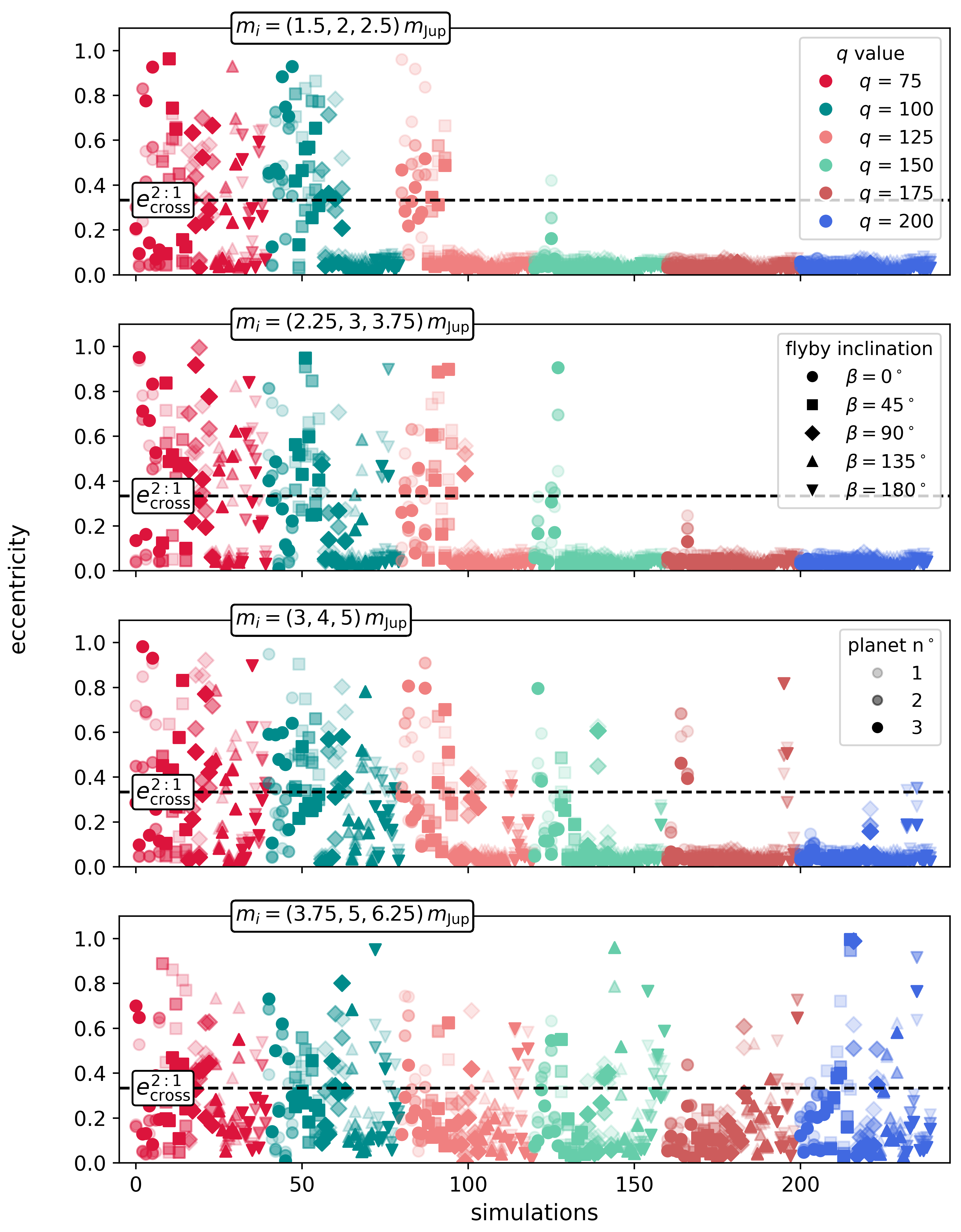} \\
    \includegraphics[width=\columnwidth]{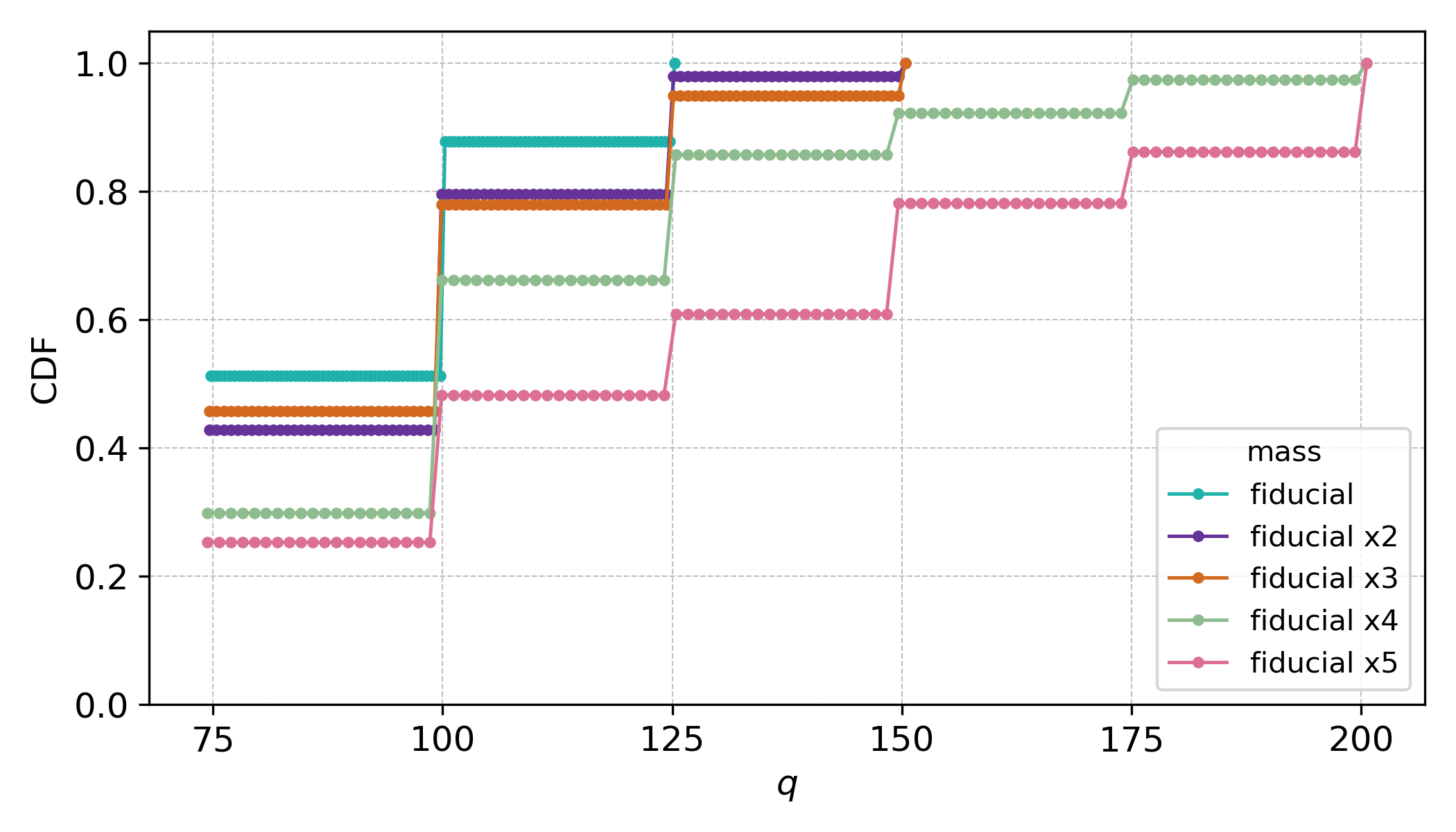} \\ 
    \caption{Four top panels show the delayed eccentricities for super-Jupiter systems starting in the 2:1-2:1 configuration. From top to bottom the set of planetary masses are $(1.5, 2, 2.5) \, m_{\rm Jup}$, $(2.25, 3, 3.75) \, m_{\rm Jup}$, $(3, 4, 5) \, m_{\rm Jup}$ and $(3.75, 5, 6.25) \, m_{\rm Jup}$. Colors and markers follow the scheme used in Figure \ref{fig:e_i_sims}. The bottom panel shows the CDF for the distances of the flyby, each curve represents a different mass configuration.}%
    \label{fig:planet_mass}
\end{figure}

\begin{figure*}
    \centering    
    \includegraphics[width=\textwidth]{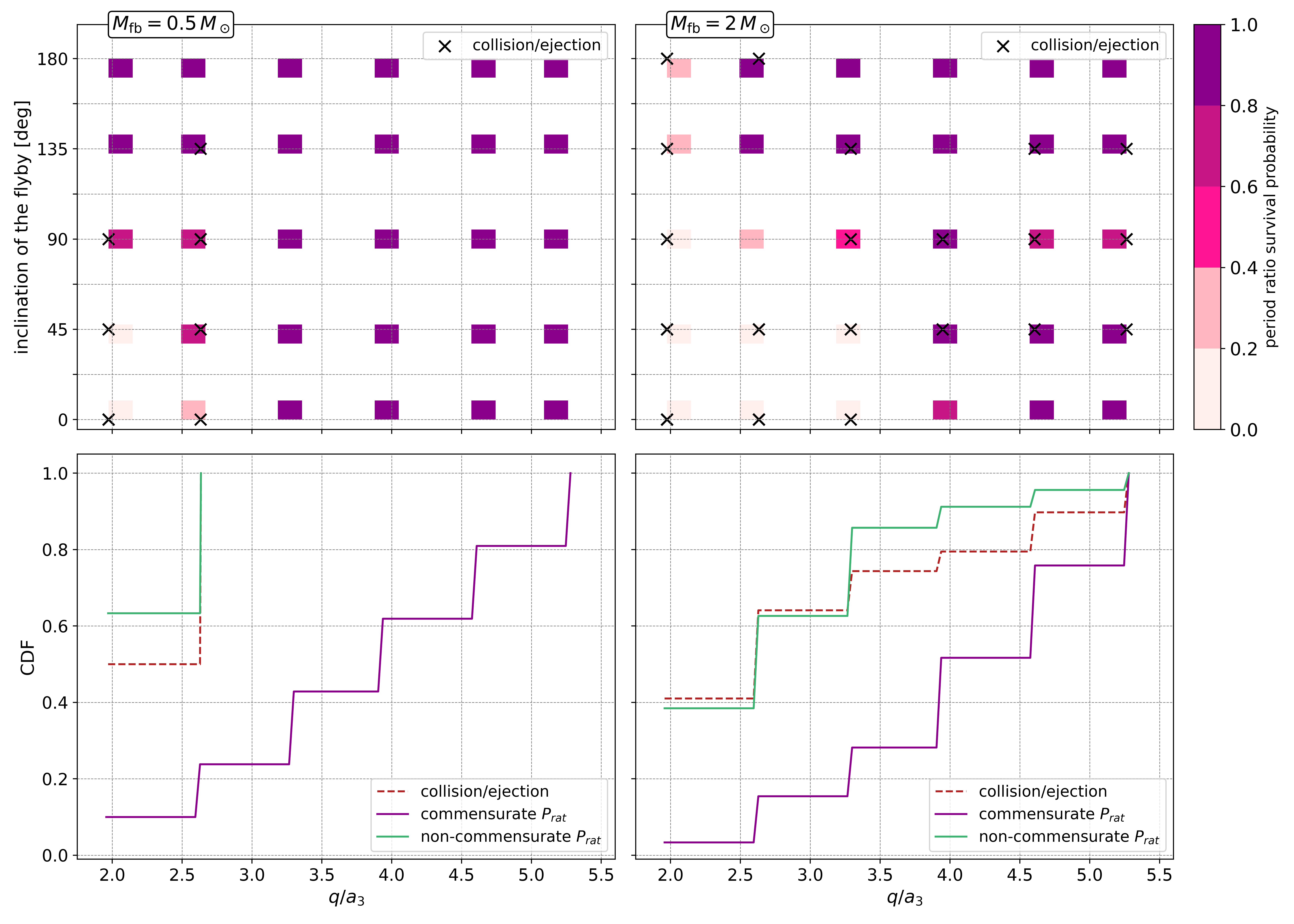}
    \caption{Survival probability of period ratios of a three-planet system with masses $(0.75, 1, 1.25) \, m_{\rm Jup}$ in 2:1-2:1 resonant chains after a star with a mass of $M_\star = 0.5 \, M_\odot$ (left) and $M_\star = 2 \, M_\odot$ (right) flies by. In the bottom panels we show the cumulative distribution for the possible outcomes of the simulations. }
    \label{fig:Mstar}
\end{figure*}

Finally, we perform a last set of simulations considering higher mass systems, keeping fixed mass ratios between the planets. For the rest of the Section, we continue the analysis and comparisons for the 2:1-2:1 resonant chain, since we are mostly interested in giant planet systems that may survive the stellar flyby, which is not the case of the 3:2-3:2 configuration in this mass range.

We explore scenarios where the fiducial planetary masses are scaled by factors of 2, 3, 4, and 5, resulting in systems with $(1.5, 2, 2.5) \, m_{\rm Jup}$, $(2.25, 3, 3.75) \, m_{\rm Jup}$, $(3, 4, 5) \, m_{\rm Jup}$ and $(3.75, 5, 6.25) \, m_{\rm Jup}$. The effects of the delayed response of these higher-mass planets to the stellar flyby are shown in Figure \ref{fig:planet_mass}, where we present the eccentricities (top four panels) and the CDF of the eccentricity crossing orbit with respect to $q$ (bottom panel) for each scenario.

As the mass of the planets increases, the gravitational influence of a stellar flyby causes greater destabilization within the planetary systems. This increase in planetary mass amplifies the perturbative effects, leading to more pronounced interactions among the planets. This increase in mass also results in a gradual growth in eccentricities for more distant flybys, reflected in the monotonic increase of the CDF (bottom panel of Figure \ref{fig:planet_mass}). Each curve represents a distinct mass configuration, illustrating how the survival probability varies with the closest approach distance.

Furthermore, as planetary masses increase, their dynamical spacing decreases (measured in mutual Hill radii), impacting the system's overall stability. For planets with masses approximately four times that of Jupiter, their Hill radius becomes comparable to that of Jupiter-mass planets in the 3:2-3:2 resonance. Consequently, the dynamical behavior of high-mass planets in the 2:1-2:1 resonance begins to resemble that of the 3:2-3:2 systems, shown in Figure \ref{fig:e_i_sims}. This is in agreement with the expectation that smaller dynamical spacings are an indicator of instability in compact configurations \citep{1996Icar..119..261C}.

As in the cases analyzed before, retrograde distant flybys increases the probability of the system retaining constant period ratios. In systems that do not cross the critical eccentricity threshold $e_{\rm cross}$ after the delayed integrations, the 2- and 3-planet resonant angles circulate, while the systems maintain constant period ratios without becoming unstable. Finally, in higher-mass systems, the difference between the immediate and delayed response to the passing star is significantly more pronounced, similar to what is observed in the 3:2-3:2 resonant chain.

\subsection{The effect of the mass of the fly-by star} \label{sect:star_mass}

In the previous section we showed that a $1\,M_\odot$ star approaching at distances of hundreds of au's can break the resonant chains. Since most stellar encounters occur between unequal mass stars  \citep{2013A&A...549A..82P,2018MNRAS.478.2700W}, we build on the fiducial case and vary the mass of the flying-by star, considering two different masses for the approaching star: $M_{\rm fb}=0.5$ and $M_{\rm fb}=2\, M_\odot$. A similar approach is used for disks in \citet{Cuello+2019}, as noted in their Appendix~D. For the planetary system, again we only focus on the case of the 2:1-2:1 resonant chain to analyze the overall impact.

The results are shown in Figure \ref{fig:Mstar}. As we established in the previous Sections, for the 2:1-2:1 resonance, the destabilization caused by the flyby is immediate, and for that reason we only show the results without the long-term additional integrations. Our simulations indicate that the larger the mass of the approaching star, the greater its destabilizing effect on the planets, in agreement with \citet{2011MNRAS.411..859M}. Moreover, for a more massive star, collisions with the central star and ejections from the system still occur even at larger relative distances, expressed as $q/a_3$. This suggests that the mass of the perturber (or the mass ratio between the host and intruding star) plays a crucial role in the dynamical evolution and stability of the resonant chains, should a flyby have occurred. However, based on stellar cluster dynamics and evolution, a $2\, M_\odot$ star approaching a Solar-mass star with planets is less likely to occur \citep{1998ApJ...508L.171L, 2013A&A...549A..82P}.

It is important to emphasize, however, that while a close approach with a 2 Solar-mass star is not expected to occur frequently, the mass ratio between the stars in the simulations resembles that of GJ 876 encountering a $1 \, M_\odot$ star. 
Conversely, when the flyby involves a star with $M_{\rm fb} = 0.5\, M_\odot$, the scenario is more similar to what HR 8799 might have experienced. This comparison allows us to place constraints on the flyby distance, assuming such an event occurred. In the left panels of Figure \ref{fig:Mstar}, we observe that if the minimum approach of the perturbing star exceeds 100 au from the planetary system, the system will most likely survive, which could explain the persistence of HR 8799-like systems, where such close encounters are less frequent. On the other hand, when a more massive star approaches the system, even distant flybys trigger collisions with the star and ejections, with fewer simulations resulting in constant resonant period ratios. A system like GJ 876 could not have survived a flyby in a wide orbit configuration.

\section{Discussion}\label{sect:conclusions}

We investigate the orbital evolution of systems with long-period planets in resonant chains, focusing on the impact of a flyby star on their stability. Our analysis considers two initial configurations: 2:1-2:1 and 3:2-3:2 resonant chains. 

The main findings can be summarized as follows:
\begin{enumerate}
    \item Stellar flybys can disrupt wide-orbit resonant chains, either by ejecting planets immediately or through delayed interactions, potentially explaining the absence of such systems.
    \item Jupiter-mass planets in 3:2-3:2 resonant chains are more easily disrupted by stellar flybys compared to 2:1-2:1 chains. 
    As anticipated, the degree of disruption of the period-ratios increases when the star flies by the system in a retrograde orbit and with decreasing distance to the planetary system.
    \item Flybys cause larger period ratio oscillations and force the resonant angles to transition out of the libration zone. Placing the outermost planet at 40~au, Jupiter-mass systems initially in a 2:1-2:1 resonant chain remain stable if the passing star remains farther than $q=125$~au. On the contrary, systems in a 3:2-3:2 configuration already become unstable for distant flybys ($q \sim 900$~au) since they are at the brink of instability due to their more compact nature.
    \item More massive planets experience disruptions across a wider range of flyby parameters, though increasing the planetary mass five times yields similar outcomes. Sub-Saturn mass planets maintain stable period ratios in both configurations, provided the flyby star's mass remains constant.
    \item For planetary masses up to that of Jupiter, the 2:1-2:1 chain responds instantaneously to flybys, whereas the 3:2-3:2 chain becomes unstable over time due to planet-planet interactions. The response of super-Jupiters to a stellar flyby in the 2:1-2:1 chain closely resembles the behavior of Jupiter-mass planets in the more compact 3:2-3:2 configuration.
    \item The disruption of period-ratios commensurabilities is greater when a more massive star flies by the planetary system.
    \item Jupiter-mass systems in wide 3:2-3:2 resonant chains are unlikely to survive stellar flybys, while lower-mass systems in wide orbits could remain stable.
\end{enumerate}

Going back to the original question, it is possible that planetary systems originally formed in wide orbit resonant chains but were later disrupted by close encounters with passing stars. Such interactions could have destabilized the resonant configuration, resulting in the systems we observe today, where these chains no longer exist. If a system originally formed in a 3:2-3:2 resonance, disruptions seem more likely compared to those in a 2:1-2:1 configuration. In the case of HR~8799, the required $q/a_3$ value for disruption is sufficiently small to make this scenario unlikely.

One limitation of this study is the relatively narrow range of parameters explored. In our simulations, planetary systems orbit stars with fixed masses, unlike the variability observed in real systems. Varying the mass of the central star leads to differences in central potential, influencing how gravitationally bound the planets are to their host stars. These variations introduce complexities beyond the resonant dynamics. Future work could broaden the parameter space by considering a wider range of stellar and planetary masses, as well as orbital configurations. It is also important to note that while we propose this parametric study, the most expected outcomes are those resembling $\beta =45^\circ$ or $135^\circ$. Stars passing with exact $\beta$ values of $0^\circ$, $90^\circ$, or $180^\circ$ are less likely to happen. 
Additionally, impulsive flybys interact with the planets for a shorter duration and may affect their orbits differently. However, such encounters are typically penetrating or highly catastrophic, likely leading to the complete disruption of the planetary system.

Finally, much remains to be explored concerning the influence of stellar flybys on the population of both detected and yet-to-be-detected exoplanets. In a resonant system, the associated angles are theoretically expected to librate. However, our simulations show that following a stellar flyby, these angles circulate instead, even as the period ratios remain constant. Thus, finding planets in such configurations --- with stable period ratios but circulating angles --- could potentially indicate that a past flyby event perturbed the planetary system.

\section*{Acknowledgments}
The authors thank the anonymous reviewer for their report that helped improve our manuscript. CC acknowledges support from Agencia Nacional de Investigación y Desarrollo (ANID) through FONDECYT post-doctoral grant n$^\circ$3230283. This project has received funding from the European Research Council (ERC) under the European Union Horizon Europe programme (grant agreement No. 101042275, project Stellar-MADE).  CP acknowledges support from ANID BASAL project FB210003 and FONDECYT Regular grant 1210425. CC and CP gratefully acknowledge support from the John and A-Lan Reynolds Faculty Research Fund.

\bibliographystyle{aa}
\bibliography{flybys}{}

\end{document}